\documentclass{article}
\usepackage{latexsym}
\usepackage{amssymb}
\usepackage{amsmath}
\usepackage{graphicx}

\begin{document}

\title{Abstract mathematical treatment of relativistic phenomena}
\author{T.~Matolcsi\thanks
{Department of Applied Analysis, E\"otv\"os University, P\'azm\'any P. s\'et\'any 1C., H--1117 Budapest, Hungary. Supported by OTKA-T 048489.},~
M.~Matolcsi\thanks
{Alfr\'ed R\'enyi Institute of Mathematics, Hungarian Academy of Sciences, Re\'altanoda utca 13--15., H--1053, Budapest, Hungary. Supported by OTKA-F 049457, T 049301, T 047276.} ~and~
T.~Tasn\'adi\thanks
{Department of Solid State Physics, E\"otv\"os University, P\'azm\'any P. s\'et\'any 1A., H--1117 Budapest, Hungary. Supported by OTKA-F 43749 }}
\maketitle

\newcommand\uu{\mathbf u}
\newcommand\Hh{\mathbf H}
\newcommand\Eu{\mathbf E_\uu}
\newcommand\h{\mathbf h}
\newcommand\x{\mathbf x}
\newcommand\y{\mathbf y}
\newcommand\I{\mathbf I} \newcommand\R{\mathbf R}
\newcommand\vv{\mathbf v}
\newcommand\M{\mathbf M}
\newcommand\V{\mathbf V}
\renewcommand\t{\mathbf t}
\newcommand\Pp{\mathbf P}
\newcommand\s{\mathbf s}
\newcommand\si{\sigma}
\newcommand\be{\begin{equation}}
\newcommand\U{\mathbf U}
\newcommand\q{\mathbf q}
\newcommand\Om{\boldsymbol\Omega}
\newcommand\E{\mathbf E}
\newcommand\z{\mathbf z}
\newcommand\zz{\hat{\mathbf z}}
\newcommand\A{\mathbf A}
\newcommand\uc{\uu_c}
\newcommand\as{\mathbf a} \newcommand\D{\mathrm D}
\newcommand\al{\alpha}
\newcommand\hal{\hat\alpha}
\newcommand\bb{\beta}
\newcommand\hbb{\hat\beta}
\newcommand\dd{\mathbf d}
\newcommand\kk{\mathbf k}
\newcommand\RR{\mathbb R}
\newcommand\F{\mathcal F}
\newcommand\De{\Delta}

\begin{abstract} 
This preprint concerns a mathematically rigorous treatment of an interesting physical phenomenon in relativity theory. We would like to draw the reader's attention particularly to the abstract mathematical formalism of relativity (which was developed in full detail in \cite{matolcsi2}). This treatment allows all mathematically oriented readers to understand relativity without feeling the awkward ambiguities that are so common after reading a standard text on relativity. 

In the extensive literature dealing with the relativistic phenomenon of Thomas rotation several methods have been developed for calculating the Thomas rotation angle of a gyroscope along a circular world line. One of the most appealing methods \cite{rindler}, however, subsequently led to a contradiction in \cite{herrera} when three different Thomas rotation angles were obtained for the same circular world line. In this paper we resolve this contradiction by rigorously examining the theoretical background and the limitations of the principle of \cite{rindler}.     
\end{abstract}

\section{Introduction}

The relativistic phenomena of Thomas rotation and Thomas precession have been
treated in relativity theory, both special and general,  from various points of 
view (see e.g. \cite{costella},  \cite{fisher}, \cite{kennedy}, \cite{matolcsi},
 \cite{moller}, \cite{philpott}, \cite{rebilas}, \cite{rindler}, \cite{thomas}, \cite{ungar},     
 \cite{wilkins}). As this preprint is aimed at a mathematical readership,  we will include a short description of the appearing physical phenomena. Also, the mathematics being used does not go beyond standard facts from linear algebra, differential geometry and calculus. We firmly believe that this abstract formalism is the appropriate language of relativity, where paradoxes and confusion simply vanish, and the physical concepts become clear. {\it Readers unfamiliar with special relativity (or, this abstract form of it) should not be discouraged; it may be helpful to study \cite{matolcsi} where the formulae are less involved.}  

Unfortunately, there seems to be no
standard agreement in the literature as to the usage of the terminologies 'Thomas precession' and 'Thomas rotation'; we will adhere to the terminologies used in \cite{matolcsi}. 
We remark, that these notions also provide a possible way to put relativity to the test in 
practice (Thomas rotation is one of the relativistic effects contributing to 
the gyroscopic precession currently being measured in the Gravity Probe B 
experiment). While the results of this paper stay exclusively 
in the realm of special relativity, the appearing concepts can also be 
generalized to general relativistic spacetime models, to which the authors hope
to return in a subsequent publication. 

To grasp the essence of Thomas rotation, let us briefly describe  
this intriguing phenomenon as an analogue of the 
well known twin paradox. Consider two twins 
in an inertial frame. One of them remains in that frame for all times, 
while the other goes for a trip in spacetime, and later returns to his 
brother. It is well-known that different times have passed for the two twins: 
the traveller is younger than his brother. 
What may be surprising is that the space of the traveller when 
he arrives back, even if he experienced no torque during his journey (i.e. he thinks that his gyroscope kept its direction throughout the journey; the meaning of this must, of course, be formulated in precise terms),
will be rotated compared to the space of his brother; this is, 
in fact, the Thomas rotation. This analogy is illuminating in one 
more respect: until the traveller returns to the original frame of 
reference it makes no sense to ask  'how much younger is the traveller compared to 
his brother?' and 'by what angle is the traveller's gyroscope 
rotated compared to that of his brother?' Different observers may give different answers.
When the traveller returns to his brother, 
these questions suddenly make perfect sense, and there is an absolute answer (independent of who the observer is)
as to how much younger and how much rotated the traveller is.

Of course, an arbitrtary inertial frame can observe the brothers continuously, and can tell
at each frame-instant, what difference {\it he sees} between the
ages of the brothers. More explicitly, as it is well known, given a world line,
an arbitrary inertial frame can tell the relation between the frame's time and 
the proper time of the world line. This relation depends on the 
inertial frame: different inertial frames establish different relations.

Similarly, an arbitrary inertial frame, observing the two brothers, can tell
at each frame-instant what  difference {\it he sees} between the 
directions of the gyroscopes of the brothers. (That is, an inertial observer sees the gyroscope of the traveller 'wobble'; this is Thomas precession.) Different inertial frames
establish different relations.

This philsophy makes a clear distinction between 
Thomas rotation and Thomas precession connected to a given world line. 
On the one hand, Thomas rotation 

-- \ makes sense only for `returning' gyroscopes,

-- \ is a discrete phenomenon (i.e. it makes sense only for the (usually) discrete set of proper time instances when the gyroscope happens to be in its initial frame of reference), 
 
-- \ is an {\it absolute} notion, i.e. independent of who observes it 
(the same angle of Thomas rotation will be measured by all inertial frames 
observing the gyroscope).

On the other hand, {\it Thomas precession}
refers to the instantenous angular velocity, {\it with respect to a particular inertial 
frame}, of a gyroscope moving along an arbitrary  world line. 
Thus, Thomas precession

-- \ makes sense for arbitrary gyroscopes with respect to arbitrary inertial frames,

-- \ is a continuous phenomenon,

-- \ is a {\it relative} notion, i.e. 
the same gyroscope may show different instantaneous precessions with 
respect to different inertial frames. 

 In terms of any 
particular inertial frame one can think of Thomas rotation as the time-integral
of Thomas precessions (and while Thomas precession, as a function of time, will
differ from one inertial frame to another, its integral will always give the 
same angle: the Thomas rotation).   

Evaluating the Thomas rotation angle, even for a gyroscope moving along a 
circular orbit, can lead to lengthy calculations. In order to arrive at the 
result in the shortest possible way, diverse approaches have been developed
in the literature (see e.g. \cite{matolcsi}, \cite{philpott},  \cite{rebilas}, 
\cite{rindler}). 

One of the simplest and most appealing concepts, introduced 
in \cite{rindler}, is to relate the Thomas precession of the gyroscope to the 
angular velocity of an observer co-moving with the gyroscope (and then 
calculate the Thomas rotation angle from the precession). 
The principle in that paper is that, heuristically, if the gyroscope keeps 
direction in 
itself and the co-moving rotating observer has instantenous angular velocity 
$\Om$ then it will see the gyroscope precess with angular velocity $-\Om$, 
and when the gyroscope returns to its initial local rest frame one can evaluate 
the Thomas rotation angle from the knowledge of instantenous precessions $-\Om (t)$ along 
the way.  

However, this principle was applied subsequently in \cite{herrera} 
to three different rotating observers co-moving with the gyroscope (the existence of such different observers is not unexpected), 
and three different Thomas rotation angles were derived; an obvious 
contradiction. In fact, the correct angle was obtained for the conventional 
rotating observer, while the `Trocheris-Takeno' and the 
`modified-Trocheris-Takeno' rotating observers gave erroneous angles. It is 
therefore desirable to  examine the theoretical background of the above 
heuristic principle and see where its limitations are, i.e. what observers 
it can justifiably be applied to. In doing so, we will also introduce the 
natural concept of Foucault precession and examine its relation to the angular 
velocity of the observer and the Thomas rotation of the gyroscope.      
    
Along the way we obtain a mathematical 
criterium for the existence of the Foucault precession in the space point of a 
noninertial observer (Section \ref{ss:fuko}). Then we check this criterium 
for different rotating observers to see whether the Foucault precession
with repect to them is meaningful or not (Section \ref{fukorot}).
It turns out that the Foucault precession is only meaningful in the case of 
the conventional rotating observer, in which case its angular velocity is 
indeed the negative of the angular velocity of the observer (in accordance with the principle in \cite{rindler}). We also conjecture,
more generally,  that whenever the Foucault precession makes sense for an 
observer, it is equal to the negative of the angular velocity of the observer. 
Finally, we examine the relation of the
Foucault precession and the Thomas rotation angle (Section \ref{contra}). 
We find that even if the Foucault precession in the space point of an 
observer makes sense, a further property of the observer is necessary so 
that the Thomas rotation angle be evaluated from the knowledge of all 
instantenous Foucault precessions.    

Throughout the paper we shall use an abstract formalism of special relativity 
(see \cite{matolcsi1}, \cite{matolcsi2}). Our basic concept is that special 
relativistic spacetime has a four-dimensional affine structure, and the 
customary coordinatization (relative to some reference frame) is, in many cases,
unnecessary in the description of physical phenomena. Besides yielding 
mathematically rigorous formulae, this coordinate-free treatment of relativity 
also allows us to make clear conceptual distinction of the appearing concepts.

\section{Fundamental notions}

In this section the necessary notions and results of the special
relativistic spacetime model as a mathematical 
structure (\cite{matolcsi1}, \cite{matolcsi2}) will be recapitulated.
As the formalism slightly differs from the usual 
textbook treatments of special relativity (but only the formalism: 
our treatment is {\it mathematically equivalent} to 
the usual treatments), we will point out several relations 
between textbook formulae and those of our formalism, and also advise the 
reader to consult \cite{matolcsi2} for a more detailed account. A concise 
summary of the appearing notions is also contained in \cite{matolcsi}. The advantage of this abstract formalism is that tacit assumptions and intuitive notions (that can go wrong so easily) are ruled out; each appearing concept (starting from the very notion of observers and synchronizations) are mathematically defined. 

\subsection{Observers and synchronizations}

 Special relativistic spacetime is a four dimensional
affine space $M$ over the vector space $\M$; the spacetime
distances form an oriented one dimensional vector space $\I$,
and an arrow oriented Lorentz form $\M\times\M\to\I\otimes\I$,
$(\x,\y)\mapsto\x\cdot\y$ is given. 

An {\it absolute velocity} $\uu$ is a future directed
element of $\frac{\M}{\I}$ for which $\uu\cdot\uu=-1$ holds (absolute 
velocity corresponds to four-velocity in usual terminology).

For an absolute velocity $\uu$, we define the three-dimensional
spacelike linear subspace 
\be
\Eu:=\{\x\in\M\mid \uu\cdot\x=0\};
\end{equation}
then 
\be 
1+\uu\otimes\uu: \M\to \Eu,\quad \x\mapsto\x+\uu(\uu\cdot\x)
\end{equation}
is the projection onto $\Eu$ along $\uu$. The restriction of 
the Lorentz form onto $\Eu$ is positive definite, so $\Eu$
is a Euclidean vector space (this will correspond to the space vectors 
of an inertial observer with velocity $\uu$).

The history of a classical material point is described 
by a differentiable {\it world line function} $r:\I\to M$
such that $\dot r(\s)$ is an absolute velocity for all proper 
time values $\s$. The range of a world line function -- 
a one dimensional submanifold -- is called a {\it world line}.

An {\it observer} $\U$ is an absolute velocity valued smooth map
defined in a connected open subset of $M$. (This is just a mathematical
 definition; it may sound unfamiliar at first, but considering that 
something that an observer calls a `fixed space-point' 
is, in fact, a world line in spacetime,
this definition will make perfect `physical' sense). 
A maximal integral curve of $\U$ -- a world line -- is a 
{\it space point} of the observer, briefly a $\U$-{\it space point}; 
the set of the maximal integral curves of $\U$ is the {\it space} of 
the observer, briefly the $\U$-{\it space}.  

For every spacetime point $x$ in the domain of $\U$ there is a
unique $\U$-space point $C_\U(x)$ containing $x$. 

A {\it synchronization} or {\it simultaneity} 
is a smooth equivalence relation on a connected open subset of
$M$ such that the equivalence classes are connected three-dimensional 
smooth submanifolds (hypersurfaces) whose tangent
spaces are spacelike (a vector $\x\in\M$ is spacelike if $\x\cdot \x >0$). 

Given a synchronization $S$, an equivalence class is called
an $S$-{\it instant}; the set $I_S$ of $S$-instants is called
$S$-{\it time}.

For every world point $x$ in the domain of $S$ there is a
unique $S$-instant $\tau_S(x)$ containing $x$; moreover,
there is a unique absolute velocity value $\U_S(x)$ such that $\E_{\U_S(x)}$
is the tangent space of $\tau_S(x)$ at $x$. The smoothness of
the synchronization means that the velocity field $x\mapsto \U_S(x)$
is smooth. (Thus an observer $\U_S$ corespsonds to every synchronization 
$S$; it is worth mentioning that there are observers which do not
correspond to any synchronization.)

A {\it reference frame} is a pair $(S,\U)$, where $S$ is
a synchronization and $\U$ is an observer. We remark that there is no a priori 
relation between $\U_S$ (the velocity field corresponding to $S$) and $\U$ 
(an arbitrary observer). Let us also mention that a reference frame makes it 
possible to `coordinatize' spacetime by $S$-instants and $U$-space points. 

An observer having consant value is called {\it inertial}.
The space points of an inertial observer are parallel straight lines.
The inertial observer with absolute velocity $\uu$
establishes its standard synchronization in which the instants
are hyperplanes over the vector space $\Eu$. An inertial observer
together with its standard synchronization is called a
{\it standard inertial frame}.

\subsection{Nearly standard local synchronizations}\label{locnear}\

A non-inertial observer $\U$ has no standard synchronization.
However, for every $\U$-space point we can give a {\it 
nearly standard local synchronization}.

More generally, if $r$ is a smooth world line function, then we
define the nearly standard local synchronization due to $r$
in a neighbourhood of the range of $r$ (the world line
determined by $r$) in such a way that the instants of that 
synchronization are subsets of spacelike  hyperplanes in such
a way that the hyperplane at an arbitrary point is Lorentz
orthogonal to the tangent vector of the world line in question.
In other words, the synchronization instant corresponding to
$r(\s)$ is a part of the hyperplane $r(\s)+\E_{\dot r(\s)}$.
The implicit function theorem assures that such a synchronization
is well defined: for fixed $\s_0\in\I$ and for $x$ in a neighbourhood
of $r(\s_0)$, the relation $(x-r(\s))\cdot\dot r(\s)=0$ can be solved 
for $\s$ and the implicit function $x\mapsto\s(x)$ satisfies
\begin{equation}\label{implido}
 \frac{d\s(x)}{dx}= - \frac{\dot
r(\s(x))}{1+(x-r(\s(x)))\cdot\ddot r(\s(x))}.
\end{equation}

Note that 
\be\label{rimplido}
\frac{d\s(x)}{dx}\Big|_{x=r(\s)}=-\dot r(\s).
\end{equation}
              
As usual, the standard inertial frame with absolute velocity value
$\dot r(\s)$ is called {\it the local rest frame} corresponding to
$r(\s)$. Roughly speaking, attaching the local rest frame
to every world point in the range of $r$, we get the above described nearly 
standard local synchronization due to $r$.

The time instants of this nearly standard local 
synchronization can be identified with the proper time values of
the world line function $r$. 

\subsection{Splitting of spacetime}\label{split}

A reference frame $(S,\U)$ {\it splits} spacetime into $S$-time 
and $\U$-space which
means that the corresponding $S$-instants and $\U$-space points
are assigned to spacetime points:
\be\label{e:splM}
 M\to I_S\times E_{\U},\qquad x\mapsto (\tau_S(x),C_\U(x)).
\end{equation}

It is well known from the theory of manifolds that both
$S$-time $I_S$ and $\U$-space $E_\U$  can be endowed with a
distinguished smooth structure, according to which both
$\tau_S$ and $C_\U$, and consequently, the splitting
will be smooth. The smooth structure of $\U$-space is defined in
such a way that
given an $S$-instant $t$ --~a hypersurface in spacetime~--, every $\U$-space
point --~a world line in spacetime~-- has a neighbourhood in $\U$-space
which is diffeomorphic with an open subset of the hypersurface $t$ via
the correspondence $q\mapsto t\cap q$; the tangent map of this
diffeomorphism sends the
tangent space of $E_\U$ at $q$  into $\E_{\U_S(t\cap q)}$, the tangent
space of the hypersurface $t$ at the meeting point of $t$ and $q$.

The derivative of $C_\U$, depending on the world points, establishes a 
mapping from the spacetime vectors to the tangent space of $E_\U$:
\be\label{e:splv}
 \D C_\U(x):\M\to 
T_{C_\U(x)}(E_\U),
\end{equation}
where $T$ denotes tangent space.

\subsection{Representation of an observer space \\ by a synchronization
instant}\label{ss:repr}

We shall apply the previous considerations to the nearly standard
local synchronization $S$ due to a $\U$-line function $r$. Then

-- \ the $S$-time instants are labelled by the proper time values
$\s$ of $r$,

-- \ the $S$-instant corresponding to $\s$ is a subset of the hyperplane
$r(\s) + \E_{\dot r(\s)}$,

-- \ $\U_S(x)=\U(r(\s(x)))=\dot r(\s(x))$, where where $\s(x)$ is defined
in Subsection \ref{locnear}.

For further investigations, let us introduce the mapping 
(the `flow' defined by the observer)
\be
\I\times M\to M, \quad (\t,x)\mapsto R(\t,x)
\end{equation}
where $\t\mapsto R(\t,x)$ is the world line function of $\U$ passing through 
the world point $x$, i.e. $\R(0,x)=x$ and 
\be\label{diffegy}
\frac{\partial R(\t,x)}{\partial\t}=\U(R(\t,x)).
\end{equation}

It follows from the uniqueness of the solutions 
of the differential equation \eqref{diffegy} that $R(\t,R(\t',x))=R(\t+\t',x)$;
differentiating it with respect to $\t'$ and then putting $\t'=0$, we have
\be\label{ubolu}
\frac{\partial R(\t,x)}{\partial x}\cdot\U(x)=\U(R(\t,x)).
\end{equation}

For a given $x$ let $\t(x)$ be the proper time value of the $\U$-line 
passing through $x$  for which the $\U$-line meets the hyperplane
$r(0)+\E_{\dot r(0)}$, i.e. $\t(x)$ is defined implicitly by
$\dot r(0)\cdot\bigl(R(\t,x)-r(0)\bigr)=0$. Then the implicit function 
theorem gives us
\be
\frac{d\t(x)}{dx}=-\frac{\dot r(0)\cdot\frac{\partial R(\t,x)}{\partial x}}
{\dot r(0)\cdot \U(R(\t,x))}\Big|_{\t=\t(x)}.
\end{equation}
Note that we have $\t(r(\s))=-\s$, $R(-\s,r(\s))=r(0)$, so with the
notation
\be\label{rs}
\R(\s):=\frac{\partial R(-\s,x)}{\partial x}\Big|_{x=r(\s)}
\end{equation}
we obtain
\be
\frac{d\t(x)}{dx}\Big|_{x=r(\s)}=\dot r(0)\cdot\R(\s).
\end{equation}

It is well known from the theory of differential equations that 
$\R(\s):\M\to\M$ is a linear bijection. Moreover, \eqref{ubolu} implies
that
\be\label{rsro}
\R(\s)\dot r(\s)=\dot r(0).
\end{equation}

Now we find it convenient to introduce the notation
\be 
\Pp(\s):=1+\dot r(\s)\otimes \dot r(\s)
\end{equation}
for the Lorentz orthogonal projection onto $\E_{\dot r(\s)}$.

We infer from \eqref{rsro} that $\Pp(0)\R(\s)\Pp(\s)=\Pp(0)\R(\s)$,
thus
\be\label{as}
\A(\s):=\Pp(0)\R(\s)\Pp(\s)=\Pp(0)\R(\s)
\end{equation}
establishes a linear bijection from $\E_{\dot r(\s)}$ onto $\E_{\dot r(0)}$
and similarly,
\be\label{asi}
\A(\s)^{-1}:=\Pp(\s)\R(\s)^{-1}\Pp(0) = \Pp(\s)\R(\s)^{-1}
\end{equation}
establishes a linear bijection from $\E_{\dot r(0)}$ onto $\E_{\dot r(\s)}$
and
\be\label{aao}
\A(\s)^{-1}\A(\s)=\Pp(\s),\qquad \A(\s)\A(\s)^{-1}=\Pp(0).
\end{equation}

According to the definition of the smooth structure of $\U$-space,
a neighbourhood of the $\U$-space point given by $r$ will be
represented by an open subset of the $S$-instant (hyperplane) corresponding to
$\s=0$ via the diffeomorphism 

-- $E_\U\to r(0) + \E_{\dot r(0)}$, 
$q\mapsto \bigl(r(0)+\E_{\dot r(0)}\bigr)\cap q$.

Then it follows from our previous considerations that

-- \ the map $M\to E_\U$, $x\mapsto C_\U(x)$ is represented by $x\mapsto
R(\t(x),x)$,

-- \ the tangent space of $E_\U$ at an arbitrary point is represented by
$\E_{\dot r(0)}$ and $\D C_\U(x)$ is represented by $\frac{dR(\t(x),x)}{dx}$.

In particular, $\D C_\U(r(\s))$ is represented by
\begin{multline}\label{dcurepr}
\frac{dR(\t(x),x)}{dx}\Big|_{x=r(\s)}=
\left(\frac{\partial R(\t,x)}{\partial\t}\otimes\frac{d\t(x)}{dx} +
\frac{\partial R(\t,x)}{\partial x}\right)\Big|_{\t=-\s, x=r(\s)}=\\
=(1+\dot r(0)\otimes\dot r(0))\R(\s) =\A(\s).
\end{multline}

\begin{figure}
\center{\includegraphics[scale=0.5]{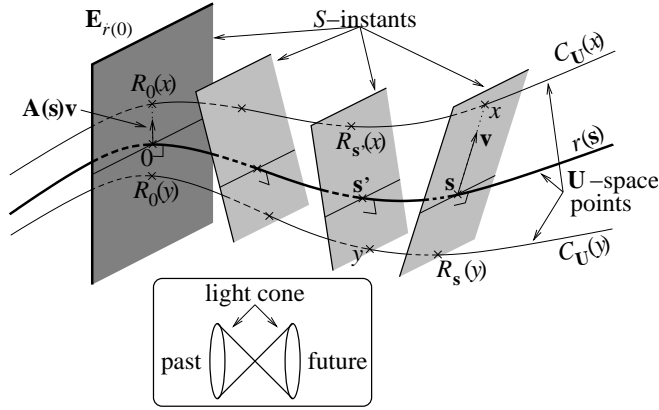}}
\caption{\label{f:osp}Representation of an observer space. For a spacetime point $x\in M$, $R(\t(x),x)=\big( r(0)+\E_{\dot r(0)}\big) \cap C_\U(x)$ is its projection along the $\U$-lines onto the $S$-instant $r(0)+\E_{\dot r(0)}$. The mapping $\A(\s)$ pulls back the vector ${\mathbf v} \in \E_{\dot r(\s)}$ `along the flow' defined by $\U$ to $\E_{\dot r(0)}$.}
\end{figure}

\subsection{Spatial metric of an observer}

The usual `spatial metric' described in coordinates by  
$\gamma_{ik}:=g_{ik} + \frac{g_{i0}g_{0k}}{-g_{00}}$
is obtained in the previous framework as follows:
we give Euclidean forms for all $\s$ (representing the 
synchronization instants) on the tangent space of all $\U$-space
points; the collection of these Euclidean forms define a Riemannian metric 
-- depending on $\s$ -- on the space of the observer 
(see \cite{matolcsi2}, Section II.9.4.). 

We will only need the local Euclidean form $\gamma_\s$ corresponding to 
the $\U$-space point described by $r$.  According to our representation of 
tangent spaces it is given on $\E_{\dot r(0)}$ and has the following 
expression:
\be\label{sthossz}
\gamma_\s(\q,\q)=|\A(\s)^{-1}\q|^2
\qquad (\q\in\E_{\dot r(0)}).
\end{equation}

\subsection{Angular velocity  of an observer}

According to the usual definition, {\it the angular velocity of 
an observer} $\U$ is 
\begin{equation}\label{e:ang}
 \Om_\U:= -\frac1{2}(1+\U\otimes\U)(\D\land\U)(1+\U\otimes\U)
\end{equation}
where $\D$ denotes differentiation, $\D\land\U:=(\D\U)^*-\D\U$ is the
antisymmetric (exterior) derivative of $\U$ (in usual coordinates:
$\D\land\U\sim \partial_iU_k-\partial_kU_i$).
$\Om_\U(x)$ is an antisymmetric linear map 
$\E_{\U(x)}\to\frac{\E_{\U(x)}}{\I}$
(in the literature, mostly the unique vector in
$\frac{\E_{\U(x)}}{\I}$ assigned to $\Om_\U(x)$ by
the Levi-Civita tensor is called the angular velocity).

The angular velocity of an observer
$\U$ refers to the change of the mutual spacetime position of
neighbouring $\U$-space points (which are world lines, maximal integral
curves of the velocity field $\U$). We call attention to the fact
that {\it the angular velocity of a single world line cannot be defined}
(see Subsection \ref{noang})

\section{Gyroscopes}

\subsection{Thomas rotation and Thomas precession}

We would like to give a precise mathematical meaning to the intuitive concept of a travelling gyroscope keeping its direction. We have to express two facts: first, the gyroscope is always spacelike according to the local rest frame and, second, the gyroscope 'keeps its direction'. We need to refer to \cite{matolcsi} for the detailed justification (in this abstract formalism) of the following standard definition: 

A gyroscopic vector 
on a world line $r$ is a pair of functions
$(r,\z):\I\to M\times\M$, where $r$ is a world line function
(the centre of the gyroscopic vector),  
$\dot r\cdot\z=0$ (the vector $\z$ is always 
spacelike according to the local rest frame), moreover,
the Fermi-Walker differential equation
\be
\dot\z = (\dot r\land\ddot r)\z=\dot r(\ddot r\cdot \z) \label{gyreq}
\end{equation}
is satisfied (which expresses the fact that $\z$ `keeps direction, 
does not rotate in itself').

\begin{figure}
\center{\includegraphics[scale=0.5]{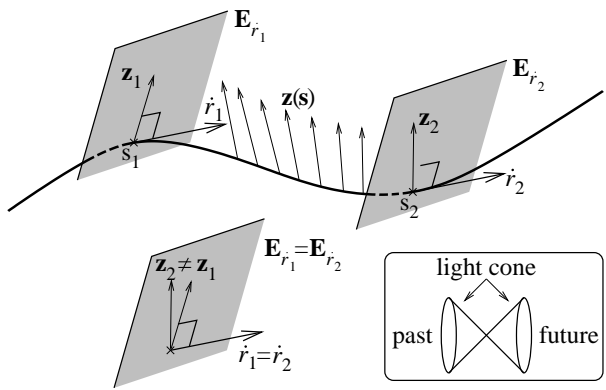}}
\caption{\label{f:t_r}Thomas rotation. At two different proper time values $\s_1$ and $\s_2$ the absolute velocities $\dot r_1 =\dot r(\s_1)$ and $\dot r_2 =\dot r(\s_2)$ are equal, so $\E_{\dot r_1} =\E_{\dot r_2}$, but the initial and final gyroscopic vectors $\z_1 =\z(\s_1) \in\E_{\dot r_1}$ and $\z_2 =\z(\s_2) \in\E_{\dot r_2}$ are different.}
\end{figure}

For proper time values  $\s_1$ and $\s_2$, the vectors $\z(\s_1)$ and
$\z(\s_2)$ are in different three-dimensional Euclidean vector spaces
unless $\dot r(\s_2)=\dot r(\s_1)$. Even if so, $\z(\s_2)\neq\z(s_1)$,
in general: the gyroscopic vector starts at $\s_1$,
tramps over diverse Euclidean spaces 'keeping its direction' in the above sense, 
and at $\s_2$ it arrives back to the starting Euclidean space
and its final direction differs from its initial direction (it
arrives rotated, see Figure~\ref{f:t_r}). This phenomenon, called 
{\it Thomas rotation}, is an absolute notion (independent of reference frames) 
and makes sense only if at least two absolute velocity values of the world 
line in question are equal.

\begin{figure}
\center{\includegraphics[scale=0.5]{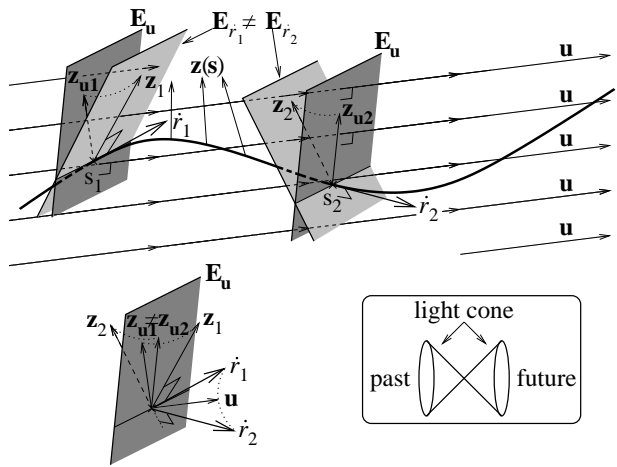}}
\caption{\label{f:t_p}Thomas precession. At every instant $\s$ the inertial observer $\uu$ boosts the gyroscopic vector $\z(\s) \in \E_{\dot r(\s)}$ to its own space $\E_\uu$, and observes that the initial vector $\z_{\mathbf u1}$ and the final vector $\z_{\mathbf u2}$ are different. In $\E_\uu$ the vector $\z_\uu$ performs a precession at an angular velocity $\Om_\uu$.}
\end{figure}

The Thomas precession, on the other hand, is a relative notion:
a standard inertial frame with velocity value $\uu$ boosts $\z$
continuously to its own space, obtaining the function $\z_\uu:I_\uu\to\Eu$
($I_\uu$ is the standard synchronization time of $\uu$) which satisfies
\be
\z_\uu'=\Om_\uu\z_\uu,
\end{equation}
where the prime denotes derivation with respect to the $\uu$-time, and
\be\label{e:omega}
\Om_\uu:= \frac{\gamma_\uu^2}{1+\gamma_\uu}\vv_\uu\land\as_\uu
\end{equation}
is the angular velocity of the precession, expressed in terms of 
the relative velocity $\vv_\uu$ and the relative acceleration $\as_\uu$
\begin{align}\label{relvel}
\vv_\uu(t) &:=\frac{\dot r(\s(t))}{-\uu\cdot\dot r(\s(t))} -\uu \\
\as_\uu(t) &:= \frac1{\big(-\uu\cdot\dot r(\s(t))\big)^2}
\left(\ddot r(\s(t))
+\frac{\dot r(\s(t))\big(\uu\cdot\ddot r(\s(t))\big)}{-\uu\cdot\dot r(\s(t))}\right)
\end{align}
of the world line. The proper time $\s(t)$ of $r$
as a function of $\uu$-time is determined by
\be
\frac{d\s(t)}{dt}= -\uu\cdot\dot r(\s(t))=
\frac1{\sqrt{1-|\vv_\uu(t)|^2}}=:\gamma_\uu(t).
\end{equation}
(For the details of the derivation of these formulae see \cite{matolcsi}.)

The standard inertial frame sees the gyroscopic vector
-- which keeps direction in itself -- precessing. Note that the same gyroscopic
vector shows different Thomas precessions with respect to different standard
inertial frames (i.e. $\Om_\uu$ really depends on $\uu$).
Note also, that different gyroscopic vectors on the same world line precess
with the same angular velocity $\Om_\uu$ with respect to the inertial frame 
$\uu$, i.e. in \eqref{e:omega} $\Om_\uu$ does not depend on $\z$.

\subsection{Foucault precession}\label{ss:fuko}

As we have seen, Thomas precession is defined with respect to {\it inertial} reference frames. 
Although the notion of Thomas precession does not seem possible to generalize to make sense with respect 
to non-inertial frames, we can introduce the natural notion of Foucault precession
with respect to observers co-moving with the gyroscope, i.e. those having the 
centre of the gyroscope as a space point. With the help of this notion we can 
investigate the validity 
of the principle described in the Introduction.

The history of a material point is perceived by a reference
frame $(S,\U)$ as a {\it motion} which is a function
assigning $\U$-space points to $S$-instants as follows.  
Let $r$ be the world line function of the material point; then
the corresponding world line meets every hypersurface $t\in
I_S$ at most in one point, thus we can give a
function $I_S\to\I$, $t\mapsto\s(t)$ such that $r(\s(t))$ is the
meeting point of the world line and the hypersurface $t$,
i.e. $\s(t)$ is the proper time of $r$ as a function of $S$-time $t$.
The unique $\U$-space point passing through the
meeting point of the world line and the hypersurface $t$
is assigned to $t$, i.e. the motion in question is
described by the function
\begin{equation}\label{mozg}
 r_{_{S,\U}}: I_S\to E_\U, \qquad t\mapsto C_\U\big(r(\s(t))\big).
\end{equation}

Then, according to the well known formulae of manifolds, the motion of a 
gyroscopic vector $(r,\z)$ with respect to the reference frame
$(S,\U)$ is described by the function 
\begin{equation}
(r_{_{S,\U}},\z_{_{S,\U}}):I_S\to T(E_\U)
\end{equation}
where $T(E_\U)$ is the tangent bundle of $E_\U$ and
\be\label{zus}
\z_{_{S,\U}}(t):= \D C_\U\bigl(r(\s(t))\bigr)\z(\s(t))
\end{equation}
(see Subsection \ref{split}).

We consider the exceptional case when
the gyroscope rests in the space of a non-inertial observer; i.e. the world 
line of the gyroscope is a space point of the observer. This 
will lead us back to giving a precise meaning to the precession principle 
described in the Introduction. 
 
The gyroscope rests in the space of a noninertial observer, keeping 
direction in itself, and the observer sees the gyroscope precessing: 
this is exactly the famous Foucault experiment, therefore we will introduce 
the terminology Foucault precession for this case. (See Figure~\ref{f:f_p}.)
The Foucault precession, a natural fact, is conceptually
different from the Thomas precession which is a counterintuitive relativistic 
phenomenon: an inertial reference frame, when observes a moving gyroscope 
which keeps direction in itself, sees a precession.

The Foucault precession in a space point of an observer $\U$ is formally 
defined as follows.

\begin{figure}
\center{\includegraphics[scale=0.5]{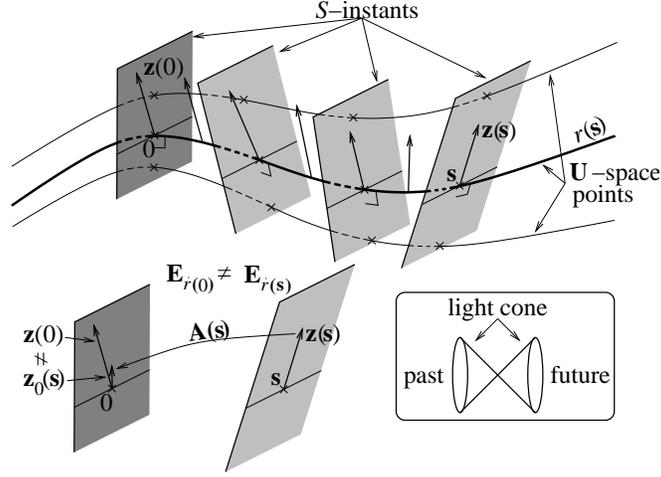}}
\caption{\label{f:f_p}Foucault precession. A noninertial `co-moving' 
observer $\U$ perceives a precession of a gyroscopic vector $\z(\s)$ whose footpoint ``rests'' at the 
space point $r$ of the observer.}
\end{figure}

Let us consider a gyroscopic vector $\z$ on the world line function $r$,
where $\dot r(\s)=\U(r(\s))$. In this case $r_{_{S,\U}}$ is constant
for an arbitrary synchronization.  Applying the nearly standard local
synchronization $S$ due to $r$, we use the formulae of Subsection \ref{ss:repr}.

Then (see \eqref{zus}, \eqref{dcurepr} and \eqref{as}) 
$\z_{_{S,\U}}(\s)$ is represented by
\be
\z_0(\s):=\A(\s)\z(\s)
\end{equation}
and we infer from \eqref{sthossz} that the length of
$\z_0(\s)$, calculated with respect to the metrics $\gamma_\s(r(\s))$, 
equals the Lorentz length of $\z(\s)$ which does not depend on $\s$.

Since $\A(\s)\dot r(\s)=0$  and $\dot\z$ is 
parallel to $\dot r$ (see \eqref{gyreq}), we have
\be\label{znul}
\dot\z_0(\s)=\dot \A(\s) \A(\s)^{-1}\z_0(\s).
\end{equation}
 Therefore our candidate for the instantenous angular velocity of the precession is $\dot \A(\s) \A(\s)^{-1}$.
Thus, the Foucault precession in the $\U$-space point given by $r$
is {\it meaningful} if and only if for all proper time values $\s$ of $r$,
the restriction of $\dot\A(\s)\A(\s)^{-1}$ onto $\E_{\dot r(0)}$ 
(and so mapping into $\frac{\E_{\dot r(0)}}{\I}$) is
antisymmetric with respect to the Euclidean form $\gamma_\s(r(\s))$,
i.e. $\bigl(\A(\s)^{-1}\q\bigr)\cdot\bigl(\A(\s)^{-1}
(\dot \A(\s)\A(\s)^{-1}\q)\bigr)=0$ for all $\q\in\E_{\dot r(0)}$. 
Equivalently, $\h\cdot\A(\s)^{-1}\dot\A(\s)\h=0$ for
all $\h \in\E_{\dot r(\s)}$ which, in turn, holds if and only if 
$\A(\s)^{-1}\dot\A(\s)\Pp(\s)$ is antisymmetric with respect to 
the Lorentz form.

Note that according to \eqref{as} and \eqref{asi}), we have
\be\label{ara}
\A(\s)^{-1}\dot\A(\s)\Pp(\s)=\Pp(\s)\R(\s)^{-1}\dot\R(\s)\Pp(\s).
\end{equation}

In section \ref{fukorot} we shall examine whether the Foucault precession 
in the space of rotating observers is meaningful or not.

Having the notion of Foucault rotation at hand we can furhter investigate 
the principle described in the introduction. Namely, we will examine the 
validity and limitations of the following assertions: 
``the angular velocity of the Foucault precession is always the 
negative of the angular velocity of the observer'' and 
``having access to the instantenous Foucault angular velocities one can 
determine the Thomas rotation angle''.

\section{Rotating observers} \label{rotobs}

\subsection{Properties of rotating observers}

Heuristically a (uniformly) rotating observer is characterized by the 
property that its space points are rotating around an inertial centre
which is the world line $o+\uu\I$, described by a specific point $o\in M$, and an 
absolute velocity $\uu \in \frac{\M}{\I}$. The rotation around the
centre, i.e. in the  spacelike hyperplane $\Eu$, is characterized by 
the angular velocity of the rotation, an antisymmetric linear map 
$0\neq\Om:\Eu\to\frac{\Eu}{\I}$ which is 
conveniently extended to the whole $\M$ in such a way that $\Om\uu=0$. 
Then at an arbitrary point $x\in M$ the velocity of the
rotation relative to the centre is proportional to 
$\Om (x-o)$, so $\U(x)$ is the linear combination of $\uu$
and $\Om (x-o)$. We restrict our attention to the case when the coefficients
in the linear combination depend only on $|\Om(x-o)|^2$ (and not on $x$). 
Thus, we accept that given positive real valued smooth functions 
$\al,\bb:\RR^+\to \RR^+$ such that
\be\label{e:a2}
\al(|\Om(x-o)|^2)^2 -\bb(|\Om(x-o)|^2)^2|\Om(x-o)|^2=1, 
\end{equation}
a corresponding rotating observer is defined as
\begin{equation}\label{e:uro}
\U(x)=\al(|\Om(x-o)|^2) \uu +\bb(|\Om(x-o)|^2))\Om(x-o).
\end{equation}
(The normalization condition \eqref{e:a2} ensures that $\U$ does  indeed 
map to the set of absolute velocities.)

For the sake of brevity, let us introduce the notation 
\be
k(x):=|\Om(x-o)|^2.
\end{equation}

Note the following special cases

1. $\al(k(x))=\bb(k(x))=\dfrac1{\sqrt{1-|\Om(x-o)|^2}}$ 
which is the conventional rotating observer (\cite{moller}, \cite{matolcsi2}),

2. $\al(k(x))=\cosh|\Om(x-o)|$,
$\bb(k(x))=\dfrac{\sinh|\Om(x-o)|}{|\Om(x-o)|}$ (the Trocheris-Takeno (TT) observer \cite{trocheris}, \cite{takeno}),

3. $\al(k(x)) = \sqrt{1+|\Om(x-o)|^2}$, $\bb=1$ (\cite{matolcsi2}),

4. $\al=\mathrm{const}>1$, $\bb(k(x))=
\dfrac{\sqrt{\al^2-1}}{|\Om(x-o)|}$.

It is worth mentioning that rotating observers usually appear
in the literature as coordinate transformations, i.e. the observers 
together with a synchronization. It is interesting to note that the 
Trocheris-Takeno transformation and the `modified Trocheris-Takeno
transformation' (MTT) (\cite{herrera2}) concern the same observer with 
different synchronizations. It is also remarkable that finding the 'right' coordinate system describing the reference frame of a rotating observer has been a minor but long-standing problem in special relativity theory (that is why the TT and MTT were introduced, after theorists were not satisfied with the conventional observer). All introduced systems have some undesireable properties. The solution to this problem is simply that there is no 'right' coordinate system: one can list the desired properties of such coordinatization, and prove that they cannot be satisfied simultaneously. 

All the $\U$-space points are circular world lines. In particular, 
the one passing through the world point $x$ is given by the function
\be \t\mapsto  o + \t\al(k(x))\uu +
e^{\t\bb((k(x))\Om}(x-o)=:R(\t,x). \label{ulinx}
\end{equation}

We shall use the following formulae:
\be\label{e:Dk}
\frac{dk(x)}{dx}=\frac{d|\Om(x-o)|^2}{dx}=-2\Om^2(x-o),
\end{equation}

\begin{align}\label{e:Dhab}
\frac{d\al(k(x))}{dx}&=-2\al'(k(x))\Om^2(x-o), \\
\frac{d\bb(k(x))}{dx}&=-2\bb'(k(x))\Om^2(x-o), 
\end{align}
where the prime denotes differentiation with respect to the real variable
of the functions. Moreover, we infer from \eqref{e:a2} that
\begin{equation}\label{aaderiv}
2\al(k)\al'(k)-2\bb(k)\bb'(k)k=\bb^2(k).
\end{equation} 

\subsection{Angular velocity of a rotating obvserver}

As a consequence of the previous formulae,
\begin{equation}
\D\U(x)=-2\bigl(\al'(k(x))\uu+\bb'(k(x))\Om(x-o)\bigr)
\otimes \Om^2(x-o)+ \bb(k(x))\Om, 
\end{equation}
so
\be
-\frac1{2}\D\land\U(x)=-\bigl(\al'(k(x))\uu+\bb'(k(x))\Om(x-o)\bigr)
\land \Om^2(x-o)+ \bb(k(x))\Om.
\end{equation}
Then, taking into account that $\Om\U(x)=-\U(x)\Om=\bb(k(x))\Om^2(x-o)$
and $\U(x)\cdot\Om^2(x-o)=0$, we can calculate the angular velocity of the
rotating observer according to 
\eqref{e:ang}: 
\be
 \Om_\U(x)=\bb\Om -\left(\left(\frac{\al\bb^2}{2}
+\al'\right)\uu +\left(\frac{\bb^3}{2}+\bb'\right)
\Om(x-o)\right)\land\Om^2(x-o) \label{omu}
\end{equation}
where, for the sake of brevity, we have written $\al$ instead of
$\al(k(x))$ etc.

\subsection{Representation of the space of a rotating observer}

Now we apply the formulae of Subsection \ref{ss:repr} to a rotating observer
by considering a fixed integral curve
\begin{align}\label{e:rs}
r(\s)&=o+\s \al_0\uu +e^{\s\bb_0\Om}\dd,&
\al_0&:=\al(|\Om\dd|^2),& \bb_0 &:=\bb(|\Om\dd|^2)
\end{align}
of the uniformly rotating observer~\eqref{e:uro}, determined by the 
initial point $o+\dd\in M$, where $\dd\in \Eu$. Then
\begin{align}\label{e:rds}
\dot r(\s) &=\al_0\uu +\bb_0e^{\s\bb_0\Om} \Om \dd,
&& \text{so}& \dot r(0) &=\al_0\uu +\bb_0 \Om \dd,
\\ \label{e:rdds}
\ddot r(\s) &= \bb^2_0e^{\s\bb_0\Om}\Om^2\dd,
&& \text{so}& \ddot r(0) &=\bb^2_0 \Om^2 \dd.
\end{align}

Next we calculate the actual form of the linear maps defined in \eqref{rs}
and \eqref{as}. The flow of the rotating observer is given 
in \eqref{ulinx}, therefore we find that 
\begin{equation}
\frac{\partial R(-\s,x)}{\partial x}= 2\s\left(\al'\uu
+\bb'e^{-\s\bb\Om}\Om(x-o)\right)\otimes\Om^2(x-o) + e^{-\s\bb\Om},
\end{equation}
where again $\al$ means $\al(k(x))$ etc. For
$x=r(\s)$ we have  $\al(k(x))=\al_0$, $\al'(k(x))=\al'(|\Om\dd|^2)=:\al'_0$ 
etc, $\Om(x-o)=e^{\s\bb_0\Om}\Om\dd$. Therefore, (see \eqref{rs})
\be 
\R(\s)= 2\s\left(\al'_0\uu
+\bb'_0\Om\dd\right)\otimes e^{\s\bb_0\Om}\Om^2\dd
+ e^{-\s\bb_0\Om}.
\end{equation}
Now using \eqref{as} and \eqref{aaderiv} leads to
\begin{equation}
\A(\s) = \Pp(0) e^{-\s\bb_0\Om}\Pp(\s)
+ \s\bigl((2\al'_0 -\al_0\bb_0^2)\uu + (2\bb'_0 -
\bb_0^3)\Om\dd\bigr)\otimes e^{\s\bb_0\Om}\Om^2\dd.
\label{derH}
\end{equation}

\subsection{Foucault precession in a space point \\ of a rotating observer}\label{fukorot}

Now let us examine whether the Foucault precession in a space point of a
rotating observer $\U$ is meaningful or not. We shall apply the formulae of
Subsection \ref{ss:fuko} for the world line function  $r$ given by
\eqref{e:rs}. The meaningfulness of the Foucault precession requires that 
$\A(\s)^{-1} \dot\A(\s)\Pp(\s)=\Pp(\s)\R(\s)^{-1}\dot\R(\s)\Pp(\s)$ be 
antisymmetric with respect to the Lorentz form. 

First we consider the case $\s=0$; then 
\be
\dot \R(0)= -\bb_0\Om +2(\al_0'\uu+\bb_0'\Om\dd)\otimes\Om^2\dd.
\end{equation}
Consequently, using \eqref{aaderiv} and $\R(0)=1$, we obtain
\begin{multline}
\Pp(0)\R(0)^{-1}\dot\R(0)\Pp(0)= \\
=-\bb_0\Pp(0)\Om\Pp(0) +
((2\al'_0 -\al_0\bb_0^2)\uu + (2\bb'_0 -\bb_0^3)\Om\dd)\otimes
\Om^2\dd.
\end{multline}

The expression on the right hand side is antisymmetric if and 
only if the second term is antisymmetric, i.e. with an abbreviated notation
$(a\uu+b\Om\dd)\otimes\Om^2\dd=-\Om^2\dd\otimes(a\uu+b\Om\dd)$.
Applying both sides to the vectors $\uu$ and $\Om\dd$, we see that
$a$ and $b$ are necessarily zero:
\begin{equation}\label{alfbet}
 2\al'_0=\al_0\bb_0^2 \qquad 2\bb'_0=\bb_0^3.
\end{equation}

If this holds, then 
\begin{equation}
\A(\s)=\Pp(0)e^{-\s\bb_0\Om}\Pp(\s) = \Pp(0)e^{-\s\bb_0\Om} =
e^{-\s\bb_0\Om}\Pp(\s),
\end{equation}
\be
\A(\s)^{-1}=\Pp(\s)e^{\s\bb_0\Om}\Pp(0) =e^{\s\bb_0\Om}\Pp(0)=
\Pp(\s)e^{\s\bb_0\Om}
\end{equation}
and $\dot\A(\s)=-\bb_0\Pp(0)\Om e^{-\s\bb_0\Om}$. Therefore,
\be\label{asasps}
\A(\s)^{-1}\dot\A(\s)\Pp(\s)=-\bb_0\Pp(\s)\Om\Pp(\s)
\end{equation}
which is evidently Lorentz antisymmetric for all $\s$. Thus, the Foucault 
precession in a spacepoint of a rotating observer is
meaningful if and only if \eqref{alfbet} is satisfied.

Since $\al_0=\al(|\Om\dd|^2)$ etc., and $\dd$ can be arbitrary, 
the  equalities in \eqref{alfbet} hold for all real variables of the 
functions, i.e. we have the differential equations
\begin{equation}\label{alfabeta}
 2\al'=\al\bb^2 \qquad 2\bb'=\bb^3.
\end{equation}

We can solve the second equation for $\bb$,  and then
taking into account (\ref{e:a2}), we find that
there is a positive constant $h$ such that
\begin{equation}
 \al(k(x))=\frac1{\sqrt{1-h^2|\Om(x-o)|^2}}, \qquad
\bb(k(x))=\frac{h}{\sqrt{1-h^2|\Om(x-o)|^2}}. \end{equation}

Therefore, we conclue that the Foucault precession in the space points of 
the rotating observers 2, 3 and 4 listed in Section \ref{rotobs} is 
not meaningful. 

The Foucault precesion is meaningful for the rotating observer 1 (for $h=1$).
Then the angular velocity of the Foucault precession at the observer
space point given by $o$ and $\dd$ is (see \eqref{asasps})
\be
\A(0)^{-1}\dot \A(0)=-\bb_0\Om + \bb_0^2(\al_0\uu +\bb_0\Om\dd)\land
\Om^2\dd.
\end{equation}
Since $o$ is an arbitrary  world point of the central world
line of the observer and $\dd$ is an arbitrary vector in $\Eu$,
we get the angular velocity of the Foucault precession 
at the world point $x$ by replacing $\al_0$ etc. with $\al(k(x))$ etc.:
\be\label{angvel}
-\bb(k(x))\Om + \bb(k(x))^2\bigl(\al(k(x))\uu + \bb(k(x))\Om(x-o)\bigr)\land
\Om^2(x-o)
\end{equation}
which is opposite to the angular velocity of the observer
at $x$ (see \eqref{omu}).

\section{Counterexamples}

\subsection{A single world line has no angular velocity}\label{noang}

One can be tempted to say that the circular world line \eqref{e:rs} 
has angular velocity  $\bb_0\Om$. 
The angular velocity of a single world line, however, cannot be defined: 
we shall show that the same world line can be a space point of 
different observers with different angular velocities.

Let $r$ be an arbitrary world line function.
Let $\s\mapsto\Hh(\s)$ be a continuously differentiable map such that 
$\Hh(\s)$ is Lorentz transformation for which $\Hh(\s)\dot r(\s)=\dot r(0)$
holds. ($\Hh(\s)$ can be for example the Lorentz boost from $\dot r(\s)$ to
$\dot r(0)$ or the Fermi-Walker transport along $r$ from $\s$ to
$0$.) Given such a family $\Hh(\s)$, and an antisymmetric linear map 
$\Gamma:\M\to\frac{\M}{\I}$ for which $\Gamma\cdot\dot r(0)=0$, the 
associated family $\Hh_\Gamma (\s):=e^{\s\Gamma}\Hh(\s)$ is another good 
choice, so we have some freedom when choosing $\Hh(\s)$. 

Taking the nearly standard synchronization instant $\s(x)$  of the 
world point $x$ (in a neighbourhood of the world line) determined 
by \eqref{implido} and putting
\be
\V(x):= \dot r(\s(x)) - 
\Hh(\s(x))^{-1}\dot\Hh(\s(x))\bigl(x-r(\s(x))\bigr),
\end{equation}
we define the observer
\be\label{ellenmf}
\U(x):=\frac{\V(x)}{|\V(x)|}
\end{equation}
where, of course, $|\V|=\sqrt{-\V\cdot\V}$.

Then $\D\U=\frac{\D\V}{|\V|} +\frac{\V\otimes(\D\V)\V}
{|\V|^3}$, and  
$\V(r(\s))=\dot r(\s)$, $|\V(r(\s)|=1$,
$\D\V(r(\s))= \ddot r(\s)\otimes\dot r(\s) - \Hh(\s)^{-1}\dot \Hh(\s)
(1 +\dot r(\s)\otimes\dot r(\s))$ and $(\D\V(r(\s))V(r(\s)=-\ddot r(\s)$. 
Since $\Hh(\s)^{-1}\dot \Hh(\s)$ is antisymmetric, we find that the 
angular velocity of the observer $\U$ (see \eqref{e:ang}) at $r(\s)$ is
\be\label{angobs}
-\Pp(\s)\Hh(\s)^{-1}\dot \Hh(\s)\Pp(\s).
\end{equation}

In particular, if $\Hh(\s)$ is the Fermi-Walker transport along $r$
from $\s$ to $0$, then $\Hh(\s)^{-1}\dot\Hh(\s)=\dot r(\s)\land\ddot r(\s)$ 
and the angular velocity of the observer at $r(\s)$ is zero. 

We emphasize what this means: {\it the properties of a world line are 
in no relation with the angular velocity of an observer 
having the world line as a space point}. 

Therefore, the circular world line \eqref{e:rs} can be
the space point of several observers having different angular velocities; 
in partiular, it can be the space point of an observer whose angular velocity 
at the world points of that circular world line is zero. 

\subsection{Thomas rotation versus Foucault precession}\label{contra}

Now we investigate the relation of the Foucault precession and the angle of 
Thomas rotation.
 
First we demonstrate that, for any choice of $\Hh(\s )$, the Foucault 
precession of the observer \eqref{ellenmf} in the space point
given by $r$ is meaningful.

It is easy to see that the function $\rho(\s):=r(\s) +\Hh(\s)^{-1}\h$ for 
$\h\in\E_{\dot r(0)}$ satisfies $\dot\rho(\s)=\V(\rho(\s))$.
Thus the range of $\rho$ is a $\U$-line ($\rho$ parameterizes
a $\U$-line by the proper time of $r$). As a consequence, using the
notations of Subsection \ref{ss:repr}, for an arbitrary world point
$x$, taking $\h:=R(\t(x),x)-r(0)$ and $\s:=\s(x)$, we get 
$x= r(\s(x)) +\Hh(\s(x))^{-1}(R(\t(x),x)-r(0))$, i.e.
\be
R(\t(x),x)=r(0) +\Hh(\s(x))(x-r(\s(x))
\end{equation}
from which we obtain by \eqref{rimplido} that
\be
\A(\s)=\frac{d R(\t(x),x)}{dx}\Big|_{x=r(\s)}=\Hh(\s)\Pp(\s).
\end{equation}

Then it follows from $\Hh(\s)\dot r(\s)=\dot r(0)$ that
\be
\A(\s)=\Hh(\s)\Pp(\s)=\Pp(0)\Hh(\s)\Pp(\s)= \Pp(0)\Hh(\s)
\end{equation}
and
\be \label{asih}
\A(\s)^{-1}=\Hh(\s)^{-1}\Pp(0)=\Pp(\s)\Hh(\s)^{-1}\Pp(0)=
\Pp(\s)\Hh(\s)^{-1}.
\end{equation}
Consequently,
\be\label{anti}
\A(\s)^{-1}\dot\A(\s)\Pp(\s)=
\Pp(\s)\Hh(\s)^{-1}\dot\Hh(\s)\Pp(\s),
\end{equation}
which is evidently Lorentz antisymmetric.

We obtained that the Foucault precession of the observer \eqref{ellenmf} in the
space point given by $r$ is meaningful and its angular 
velocity at the proper time value $\s$ equals the negative 
of the angular velocity of the observer \eqref{angobs}. The authors conjecture 
that the following general statement is true: whenever the Foucault precession 
in a space point of an observer is meaningful, it equals the negative of the 
angular velocity of the observer. We also conjecture that the Foucault conjecture is meaningful if and only if the observer is (locally) rigid. (We do not formally define rigidity here, but heuristically it simply means that the distance between any two space-points of the observer does not change in time, i.e. its spatial metric is time-independent, at least locally.) These two general statements would justify the first part of the principle of \cite{rindler} for rigid observers. We note, however, that in the calculations above only some very special cases were considered.

Note also that the gyroscope whose centre is the world line given by $r$
shows different Foucault precessions with respect to different
observers (the same gyroscope, whose centre rests in different
observers' spaces, shows different Foucalt precessions with
respect to the observers; this is not surprising, having seen that 
different observers, having $r$ as a space point, may have different 
angular velocities).

Now, if the Foucault precession is meaningful, then the time-integral of the
Foucault precession results in a finite angle and we can investigate the
connection between such a Foucault angle and the angle of Thomas rotation.
By some examples we will show that, in general, they differ from each other.

First, let us consider a trivial example. Take the world line function given in 
\eqref{e:rs} with $\dd=0$ (a space point of the axis of rotation). 
Then a gyroscopic vector $\z$ on the inertial world line function
$\s\mapsto o+\uu\s$ is constant, so the Thomas rotation is meaningful 
for every proper time value $\s$ of $r$ and equals the identity map 
(no rotation occurs). On the other hand, the Foucault precession in that space
point of the conventional rotating observer has angular velocity $-\Om$, so 
$e^{-\s\Om}$ is the rotation arising from the Foucault precession.

More generally, we have shown that the world line function given by \eqref{e:rs}
with $\dd\neq0$ can be a space point of different observers \eqref{ellenmf} 
with different meaningful Foucault precessions.

According to \eqref{asih}, the Euclidean form \eqref{sthossz}
is the restriction of the Lorentz form for all $\s$. Thus
it is meaningful that, after 
the time period $\s$, the Foucualt precession results in an angle 
whose cosine is $\frac{\z_0(0)\cdot\z_0(\s)}{|\z_0(0)|^2}$,
$\z_0$ being the solution of the differential equation \eqref{znul}.

For the conventional rotating observer
\begin{multline}
\dot \A(\s)\A(\s)^{-1}=
-\bb_0\Om + \bb_0^2\bigl(\al_0\uu+\bb_0\Om\dd\bigr)\land\Om^2\dd
= \\ =-\bb_0\Om + \dot r(0)\land\ddot r(0)=:\Om_r
\end{multline}
is independent of $\s$, thus $\z_0(\s)=e^{\s\Om_r}\z_0$, therefore
the angle in question is $\s|\Om_r|$ where
$|\Om_r|=-\frac1{2}\mathrm{Tr}(\Om_r^2)$.
We find (recall that now $\al=\bb$) that
\be
|\Om_r|=\frac{\bb_0\omega}{\sqrt{1-\omega^2|\dd|^2}};
\end{equation}
as a consequence, the Foucault angle for $\s=\frac{2\pi}{\bb_0\omega}$ 
 -- after a whole revolution -- equals the angle of the Thomas rotation 
(see \cite{rindler}, \cite{matolcsi}).

However, for other rotating observers other Foucault angles are obtained 
after a whole revolution.
Namely, let $\Gamma$ be a Lorentz antisymmetric map for which $\Gamma \dot r(0)
=0$ holds and $\Hh(\s):=e^{\s\Gamma}e^{-\s\bb_0\Om}$. Then
$\z_0(\s)=e^{\s\Gamma}e^{\s\Om_r}\z_0$ and the Foucault angles
are different for different $\Gamma$-s.

The reason behind these examples seem to be that although the world line $r$ 
returns to its initial local rest frame, i.e. $\dot r(0)=\dot r(\s_1)$, 
the velocity field given by the observer does not (except 
for the one spacepoint given by $r$). On the other hand, it is obvious that 
if the mapping $\A (\s_1 )$ is the identity (which happens to be the case
for the conventional rotating observer), then the measured angle will indeed 
give the Thomas rotation angle.

\section{Conclusion}

In conclusion we can say that there are some limitations in applying the 
principle of relating the Thomas rotation angle to the angular velocity of a 
co-moving observer. 

First, the angular velocity of the observer does not always equal the negative 
of the angular velocity of the Foucault precession, because the latter might 
not even be meaningful. (On the other hand, in our examples whenever the 
Foucault precession made sense, it was also equal to the negative of the 
angular velocity of the observer. We conjecture that such is the 
case for all locally rigid observers.)

Secondly, even if the Foucault precession is meaningful, the Foucault angle
after a whole revolution will not necessarily give the angle of Thomas 
rotation. 

Thus we can see why the correct Thomas rotation angle emerged for the 
conventional rotation observer, and what went wrong in the cases of 
Trocheris-Takeno and modified-Trocheris-Takeno observers in \cite{herrera}.

\end{document}